\documentclass[letterpaper,twocolumn,10pt]{article}
\usepackage[textwidth=7in,textheight=9in,centering]{geometry}
\usepackage[T1]{fontenc}
\usepackage{mathptmx}
\usepackage{graphicx}
\usepackage{amsmath,amssymb,amsthm}
\usepackage{booktabs,tabularx,array,multirow}
\usepackage{url}
\usepackage{xurl}
\usepackage{microtype}
\usepackage{cite}
\usepackage{tikz}
\usepackage{hyperref}
\usetikzlibrary{arrows.meta,positioning,fit,backgrounds,calc}
\hypersetup{colorlinks=true,allcolors=blue!65!black,
  pdftitle={Signed, Valid, and Unauthorized: Authority Closure at the Trust Boundary Between Agent Commerce Protocols and Crypto Wallets},
  pdfauthor={Shengwei You, Aditya Joshi, Andrey Kuehlkamp, and Jarek Nabrzyski},
  pdfsubject={Author preprint}}
\newcommand{\NCases}{2,130}
\newcommand{\NViol}{1,680}
\newcommand{\NBenign}{450}
\newcommand{\NPrograms}{62,979}
\newcommand{\NFuzz}{42,600}
\newcommand{\NProgAccepted}{13,278}
\newcommand{\FBUnauth}{265}

\newcommand{\ClosureFA}{0}
\newcommand{\ClosureFR}{0}
\newcommand{\SimFA}{840}
\newcommand{\FactFA}{570}
\newcommand{\FactTFA}{295}
\newcommand{\PolicyFA}{646}

\newcommand{\PolicyFR}{108}

\newcommand{\ReasonExact}{1,680}
\newcommand{\TargetedFound}{1,680}

\newcommand{\RealLossCases}{1,391}

\newcommand{\LatMax}{12}
\newcommand{\LatMin}{3}
\newcommand{\LatPMax}{20}
\newcommand{\AblNoRegFR}{150}
\newcommand{\AblNoCavFR}{60}
\newcommand{\AblNoTimeFA}{300}
\newcommand{\AblNoRecFA}{360}
\newcommand{\AblNoAmtFA}{434}
\newcommand{\TraceCase}{1020}
\newcommand{\TraceValue}{285.09}
\newcommand{\TraceDays}{31.9}
\newcommand{\TraceDeadlineH}{30.8}
\newcommand{\TraceExpiryH}{46.4}
\newcommand{\SimStateFA}{630}
\newcommand{\TypeFA}{175}

\newcommand{\XApp}{780}
\newcommand{\XHonFAApp}{120}
\newcommand{\XMerFAApp}{690}
\newcommand{\RelViol}{1,620}
\newcommand{\RelFA}{0}
\newcommand{\RelFR}{0}
\newcommand{\StrictOnly}{60}
\newcommand{\RealBeyond}{496,536}
\newcommand{\RealWithin}{195,819}
\newcommand{\MutTotal}{22}
\newcommand{\MutDecision}{15}
\newcommand{\MutAny}{21}
\newcommand{\NClasses}{25}
\newcommand{\NFamilies}{13}
\newcommand{\InvSample}{71}
\newcommand{\InvUnauth}{56}
\newcommand{\InvFound}{49}

\newcommand{\InvMissed}{7}

\newcommand{\sys}{\textsc{Closure}}
\newcommand{\M}{\mathcal{M}}
\newcommand{\Auth}{\mathrm{Auth}}
\newcommand{\sem}[1]{[\![#1]\!]}
\newtheorem{definition}{Definition}
\newtheorem{proposition}{Proposition}
\newtheorem{theorem}{Theorem}
\newtheorem{corollary}{Corollary}
\date{\small Preprint \textbullet{} 18 November 2026}
\title{\Large\bf Signed, Valid, and Unauthorized: Authority Closure at the Trust Boundary\\Between Agent Commerce Protocols and Crypto Wallets}
\author{Shengwei~You, Aditya~Joshi, Andrey~Kuehlkamp, and Jarek~Nabrzyski\\[0.4ex]
  \normalsize Department of Computer Science and Engineering\\
  \normalsize University of Notre Dame, IN 46556, USA}
\begin{document}
\maketitle
\raggedbottom

\begin{abstract}
Agent commerce protocols let software agents pay with stablecoins. A buyer approves a payee, asset, spending cap and time window, but the wallet signs an object whose effects depend on the contracts that consume it. Matching the displayed fields or simulating one payment can miss transfers that the signed object still permits. We define authority closure as the set of transfer sequences that holding the signed object can enable. A wallet object respects the mandate when every such sequence meets the buyer's constraints. We characterize when wallet objects can satisfy this property and implement Closure, a wallet-side checker covering 13 families of signed objects. In an evaluation of 2,130 generated objects and 62,979 execution programs, 265 objects with an observed violation passed both a field-and-expiry check and an address allow-list. Under the test model, Closure rejected all 1,680 objects with an observed violation and accepted the remaining 450. In two local checkouts using the x402 reference client, approved payments settled and Closure refused requests outside the mandate before signing. The median decision time was at most 12 microseconds per family.
\end{abstract}

\section{Introduction}
\label{sec:intro}
A person asks a shopping agent to buy a coffee subscription: pay \emph{Bean \& Byte} at most 25 dollars in USDC on Base, by Friday. The agent receives an HTTP 402 response from the merchant, builds a payment and asks the wallet to sign. Suppose the wallet is asked to sign an EIP-2612 permit~\cite{spec:erc-2612} that names the merchant's payout address as spender, exactly 25 USDC as value, the USDC contract on Base as verifying contract and Friday as deadline. The displayed address, amount, asset, network and deadline match the buyer's instructions. A wallet that displays fields shows the right payee, amount, token, chain and date, and a transaction simulator shows no balance change. Yet the permit creates an allowance that outlives its deadline: the deadline bounds when the permit can be submitted, and nothing bounds when the allowance can be spent. Whoever controls the merchant's key, today or after a breach next month, can move 25 USDC to any address at any later time.

This example is one instance of a general gap. Agent commerce protocols describe a payment in commerce terms. AP2 mandates name merchants, fiat amounts in minor units and expiry times~\cite{spec:ap2-spec,spec:ap2-opm-schema}; x402 payment requirements name a network, an asset, a recipient address and an amount~\cite{spec:x402-v2,spec:x402-exact-evm}; ERC-7715 lets an application request a permission from a wallet~\cite{spec:erc-7715}. The wallet signs an object in contract terms, and the contract that consumes the object fixes its meaning. Recent analyses of these protocols ask whether the payee, amount, asset and chain survive the hand-off~\cite{jiang2026formal,aviv2026mandate,louck2026whisper,li2026fiveattacks,wang2026x402}. That question is necessary. It is insufficient, because a signed object can carry the right facts and still grant more than the person approved.

\begin{figure}[t]
\centering
\begin{tikzpicture}[font=\scriptsize, >=Stealth,
  box/.style={draw, rounded corners=2pt, align=center, minimum height=7mm, inner sep=2.5pt},
  adv/.style={box, fill=red!8}, tru/.style={box, fill=green!8}]
\node[tru] (p) at (0,0) {Principal\\approves mandate $\M$};
\node[adv] (a) at (0,-1.35) {Agent};
\node[adv] (m) at (3.9,-1.35) {Merchant /\\facilitator};
\node[tru] (w) at (0,-2.75) {Wallet\\+ \sys{}};
\node[box] (c) at (3.9,-2.75) {Token, Permit2,\\proxy, delegation\\manager};
\draw[->] (p) -- node[right, align=left]{payee, cap,\\asset, expiry} (a);
\draw[->] (m) -- node[above]{402 requirements} (a);
\draw[->] (a) -- node[right]{object $\sigma$ to sign} (w);
\draw[->] (w) -- node[above]{signed $\sigma$} (c);
\draw[->, dashed, red!70!black] (m.east) to[out=-10,in=10] node[right, align=left, font=\tiny]{any holder\\of $\sigma$: any\\time, recipient,\\repetition} (c.east);
\node[draw=gray, dashed, rounded corners, fit=(w)(c), inner sep=3pt, label={[font=\scriptsize\itshape]below:boundary check: $\Auth(\sigma)\subseteq\sem{\M}$?}] {};
\end{tikzpicture}
\caption{The trust boundary. The mandate is stated in commerce terms; the wallet signs an object whose authority is fixed by contract code and exercised by whoever holds it. Red: untrusted parties; green: trusted. $\Auth(\sigma)$: every transfer sequence any holder of $\sigma$ can cause; $\sem{\M}$: the sequences the mandate allows (Section~\ref{sec:model}).}
\label{fig:seam}
\end{figure}
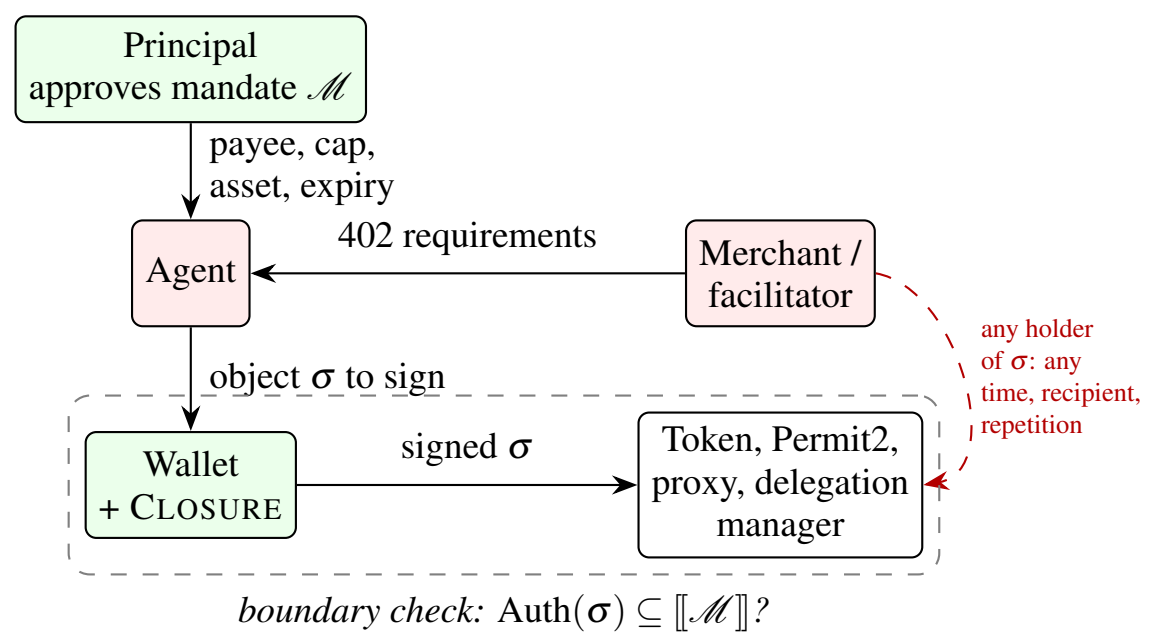

\paragraph{The property.} We define the security property on executions. A wallet object $\sigma$ \emph{refines} a mandate $\M$ if every sequence of token transfers out of the signer that an adversary holding $\sigma$ can cause, under the deployed contracts, is allowed by $\M$ (Section~\ref{sec:model}). We call the set of such sequences the \emph{authority closure} of $\sigma$. The adversary may control the agent, the merchant, the facilitator and the spender; it chooses when to submit, may replay, may split amounts and may route through approvals. Figure~\ref{fig:seam} shows where the check belongs. Three widely used checks fall short of this property, each for a structural reason. \emph{Fact binding} compares displayed fields and cannot see what the consuming contract does with them. \emph{Requirement-relative verification}, as in an x402 facilitator, checks $\sigma$ against requirements that the merchant itself supplied. \emph{Simulation} executes one element of the closure and misses authority that persists or that only an adversarial executor exercises.

\paragraph{What can refine.} The definition yields a sharp classification (Theorem~\ref{thm:char}). A wallet object can refine a mandate with a finite payee set, cap and expiry only if it \emph{names} who is paid, \emph{caps} what is consumed and has an execution window within the mandate. The time bound must hold when the transfer executes. Approvals, EIP-2612 permits, Permit2 allowances and Permit2 signature transfers without a witness let the spender pick the recipient, so under the strict definition no choice of field values makes them safe to hand to an agent. (An approval whose spender is Permit2 itself grants no direct authority and is treated separately.) ERC-3009 authorizations, the x402 Permit2 witness transfers and ERC-7710 delegations that carry an amount, a recipient pin and a time bound can refine; whether a given one does depends on its fields.

\paragraph{The checker.} \sys{} computes a footprint for a wallet object: the chain, the asset, the recipients any executor can choose, the cumulative amount and the time window in which transfers are possible. It then checks the footprint against the mandate. The footprint follows the semantics of the consuming contract: a Permit2 signature has a different footprint depending on the code of the spender it names, and an ERC-7710 delegation has the footprint left after each known caveat enforcer narrows unrestricted authority. Unknown spender code and unknown caveats receive no credit, so \sys{} fails closed.

\paragraph{The evaluation.} A footprint calculus is useful only if it matches the code it describes. We built an executable oracle from the upstream sources of Permit2~\cite{spec:permit2}, the x402 Exact and Upto Permit2 settlement proxies~\cite{spec:x402-exact-evm} and the MetaMask delegation manager with seven caveat enforcers~\cite{spec:metamask-df}, and upgrade the signer's account through EIP-7702~\cite{spec:eip-7702}. Section~\ref{sec:limitations} lists the few substitutions needed to compile offline. For each artifact the oracle executes targeted and randomized adversary programs and records every transfer out of the signer; an artifact is unauthorized if some program realizes a transfer the mandate forbids. Our corpus contains \NCases{} artifacts in \NFamilies{} families and \NClasses{} classes of correct and faulty translations. Because we wrote both the checker and the adversary library, we add two checks that do not depend on that library: Forge's invariant fuzzer, which composes operations on its own, and mutation analysis of the checker's rules.

\paragraph{Findings.} Every baseline accepts unauthorized artifacts. \FBUnauth{} unauthorized artifacts pass fact binding with an expiry check \emph{and} a Safe-style policy guard that allow-lists addresses. They include EIP-2612 permits, Permit2 signature transfers and Permit2 allowances whose displayed payee, amount, asset, chain and expiry all match the mandate, and authorizations that start before the mandate or whose total over periods or retries exceeds the cap. Simulation accepts \SimFA{} of \NViol{} unauthorized artifacts, and \SimStateFA{} when it also checks for residual allowances. Where x402 verification applies, it accepts \XHonFAApp{} of \XApp{} unauthorized artifacts with requirements derived from the mandate and \XMerFAApp{} with requirements written by the merchant. \sys{} accepted none of the \NViol{} unauthorized artifacts and rejected none of the \NBenign{} authorized ones, and for all \ReasonExact{} unauthorized artifacts the dimensions it reported (recipient, amount, time, asset, chain) equal those that execution violated.

\paragraph{Contributions.}
\begin{enumerate}
\item A definition of mandate refinement by wallet objects in terms of authority closure, three propositions showing why fact binding, requirement-relative verification and single-execution simulation do not imply it, and a characterization of the wallet objects that can refine a payee-bound mandate (Section~\ref{sec:model} and Theorem~\ref{thm:char}).
\item A footprint calculus for \NFamilies{} wallet artifact families used in agent payments, including the spender-dependent semantics of Permit2 witnesses, compositional reasoning over ERC-7710 caveats and EIP-7702 code authorizations, with a soundness theorem and an implementation as \sys{} (Section~\ref{sec:closure}).
\item An executable oracle over the upstream Permit2, x402 and delegation-framework contracts and a corpus of \NCases{} artifacts in \NClasses{} classes, with targeted, randomized and invariant-fuzzing adversaries (Section~\ref{sec:method}).
\item A measurement of which classes each deployed check admits, a differential and mutation-based validation of \sys{} against execution, and concrete requirements for protocol, wallet and permission designers (Sections~\ref{sec:results} and~\ref{sec:implications}).
\end{enumerate}
Appendix~\ref{app:openscience} describes the materials used to evaluate \sys{} and the results reported in this preprint.

\section{Background}
\label{sec:background}
\paragraph{Agent commerce protocols.} AP2 structures a purchase as signed mandates: an open mandate that the person approves on a trusted surface, a checkout mandate that binds a cart, and a closed payment mandate signed by the agent's key~\cite{spec:ap2-spec}. The open mandate in the reference implementation constrains an amount range in fiat minor units and a list of allowed payees given as merchant objects (\texttt{mandate\_tools.py}, lines 314--328 of the sample at commit \texttt{e1ea56d}); its x402 payment path then signs an ERC-3009 authorization whose recipient and value come from the payment requirements (\texttt{x402\_psp\_mcp/server.py}, lines 174--207)~\cite{spec:ap2-x402-psp}. x402 revives HTTP status 402: a resource server answers with payment requirements, the client returns a signed payment, and a \emph{facilitator}, a service that verifies and submits payments for the merchant, settles it on chain~\cite{spec:x402-v2}. Its \emph{exact} scheme on EVM chains pays with an ERC-3009 authorization, a Permit2 witness transfer or an ERC-7710 delegation; its \emph{upto} scheme lets a named facilitator settle any amount up to a signed maximum~\cite{spec:x402-exact-evm,spec:x402-exact-evm-7710}. ACP and Visa's Trusted Agent Protocol address card-based checkout and agent identity~\cite{spec:acp-dp-rfc,spec:tap-ref}; we do not study card rails.

\paragraph{Wallet artifacts.} A wallet signs either a transaction or EIP-712 typed data~\cite{spec:eip-712}. Amounts are integers in a token's \emph{base units}: one USDC is $10^6$ base units, one WETH is $10^{18}$. ERC-20 \texttt{transfer} moves value immediately; \texttt{approve} creates an \emph{allowance}, a standing right for a named spender to call \texttt{transferFrom} and move up to the approved amount to any recipient until it is revoked~\cite{spec:erc-20}. EIP-2612 \texttt{permit} creates the same allowance from a signature~\cite{spec:erc-2612}. ERC-3009 \texttt{transferWithAuthorization} moves a signed value to a signed recipient within a signed window, and \texttt{receiveWithAuthorization} additionally requires the recipient to submit~\cite{spec:erc-3009}; USDC implements both~\cite{spec:circle-fiattoken}. Permit2 is a shared contract to which users grant one standing approval; it then moves tokens under its own signatures~\cite{spec:permit2}. Its \emph{signature transfers} are single-use, and the spender chooses the recipient at execution unless a \emph{witness}, extra typed data that the signature covers, is interpreted by the spender's code to bind it. Its \emph{signature allowances} have an expiration. ERC-7710 delegations let a smart account authorize a delegate to make the account execute calls, subject to \emph{caveat enforcers}: contracts that the delegation manager calls before and after each execution and that can revert it~\cite{spec:erc-7710,spec:metamask-df}. EIP-7702 lets an ordinary account install code by signing an authorization that names an implementation contract~\cite{spec:eip-7702}. ERC-7715 is the wallet RPC through which applications request such permissions~\cite{spec:erc-7715}.

\paragraph{Where these objects arise.} Table~\ref{tab:paths} lists the agent-payment paths in current specifications and SDKs and the wallet objects they ask a wallet to sign. Two observations matter for what follows. First, the x402 exact scheme accepts ERC-7710 delegations, including multi-use ones, and verifies them ``entirely through simulation'' (\texttt{scheme\_exact\_evm.md}, lines 13 and 292). Second, the x402 gas-sponsoring extensions ask the wallet to sign an EIP-2612 permit or a raw \texttt{approve} transaction that names Permit2 as spender, for the facilitator to submit. These objects grant authority to Permit2 alone, which in turn moves tokens only under further signatures; Section~\ref{sec:closure} treats them accordingly.

\begin{table}[t]
\centering
\caption{Agent-payment paths and the wallet objects they produce (x402 at commit \texttt{0e6e8c5}; AP2 sample at \texttt{e1ea56d}). Family numbers as in Table~\ref{tab:footprint}.}
\label{tab:paths}
\footnotesize
\begin{tabularx}{\columnwidth}{@{}>{\raggedright\arraybackslash}X>{\raggedright\arraybackslash}p{2.75cm}c@{}}
\toprule
Path (source) & Wallet object & Fam.\\
\midrule
AP2 x402 PSP sample (\texttt{server.py} 174--207) & ERC-3009 transfer & 4\\
x402 exact, EIP-3009 method & ERC-3009 transfer & 4\\
x402 exact, Permit2 method & witness to Exact proxy & 7\\
x402 upto & witness to Upto proxy & 8\\
x402 exact, ERC-7710 method (spec.\ l.\,13, 292) & delegation, one-time or multi-use & 10\\
x402 EIP-2612 gas sponsoring (\texttt{client.py} 58--61) & permit, spender Permit2 & 3\\
x402 ERC-20 approval sponsoring (\texttt{client.py} 50--54) & signed \texttt{approve}(Permit2, $2^{256}{-}1$) & 2\\
ERC-7715 permission request & delegation & 10\\
EIP-7702 account upgrade & code authorization & 12\\
\bottomrule
\end{tabularx}
\end{table}

\section{Model and Security Definition}
\label{sec:model}
\paragraph{Parties and trust.} A \emph{principal} holds a signing key and a token balance and approves a mandate on a trusted surface. The agent, the merchant, the facilitator and any spender or delegate are untrusted: any of them may be compromised, prompt-injected or malicious. The adversary may control all of them, observes every signed object and chooses which transactions to submit, when, in what order and how often. The token contracts, Permit2, settlement proxies, delegation manager and caveat enforcers are trusted to behave as their source code specifies. The wallet's checker is trusted and knows the mandate. We do not consider key theft, consensus attacks, or front-running that changes no transfer out of the principal.

\begin{definition}[Mandate]
A mandate $\M=(c,a,Y,\kappa,[t_0,t_1])$ names a chain $c$, an asset contract $a$, a finite set of payee addresses $Y$, a cumulative cap $\kappa$ in the asset's base units and a time window. Its meaning $\sem{\M}$ is the set of finite transfer sequences $(c_i,a_i,\mathit{to}_i,v_i,\tau_i)_i$ out of the principal with $c_i=c$, $a_i=a$, $\mathit{to}_i\in Y$, $\tau_i\in[t_0,t_1]$ for every $i$ and $\sum_i v_i\le\kappa$.
\end{definition}

Commerce protocols carry these facts in other forms: a merchant identifier instead of an address, fiat minor units instead of base units, a currency code instead of a contract. Translating them needs trusted registries for payee addresses, assets and decimals (Section~\ref{sec:deploy}). For a single purchase, the cap is the cart total. We assume the wallet holds the translated mandate.

\begin{definition}[Authority closure and refinement]
Fix the deployed contracts and the principal's standing state $W$, for example a standing approval of Permit2. The authority closure $\Auth_W(\sigma)$ of a signed object $\sigma$ is the set of transfer sequences out of the principal that the adversary can cause in $W$ using $\sigma$ and no other signature of the principal. When $\sigma$ is a pair, the adversary receives both signatures and may submit either in any order and retry either. $\sigma$ refines $\M$, written $\sigma\sqsubseteq\M$, iff $\Auth_W(\sigma)\subseteq\sem{\M}$.
\end{definition}

The definition does not assume that the named spender, facilitator or delegate behaves honestly; a merchant whose key is stolen after checkout is the case that matters for a standing authorization. Some deployments trust the payee itself. We therefore also report a \emph{relaxed} variant in which a transfer whose recipient was chosen by an executor that is itself in $Y$ counts as paying that executor. The relaxed variant admits, for example, an allowance spent by the payee to an exchange deposit address; Section~\ref{sec:results} reports \sys{}'s decisions under both variants; all other results use the strict definition.

\paragraph{Three insufficient checks.} Let $\mathit{facts}(\sigma)$ be the payee, amount, asset, chain and expiry fields that a wallet displays for $\sigma$, and let $\mathit{Req}$ be x402 payment requirements.

\begin{proposition}[Fact binding]
There exist $\sigma$ and $\M$ with $\mathit{facts}(\sigma)$ equal to the facts of $\M$ and $\sigma\not\sqsubseteq\M$.
\end{proposition}
\noindent\emph{Witness.} The permit of Section~\ref{sec:intro}. Its spender is in $Y$, its value equals $\kappa$, its asset and chain match and its deadline is at most $t_1$. After submission the allowance permits \texttt{transferFrom} to any address at any time; our oracle realizes such a transfer 30 days after $t_1$ (Section~\ref{sec:results}). An \texttt{approve} transaction, a Permit2 signature transfer without a witness and a Permit2 allowance are further witnesses.

\begin{proposition}[Requirement-relative verification]
A verifier that accepts $\sigma$ iff $\mathit{facts}(\sigma)$ agrees with $\mathit{Req}$ provides no guarantee about $\M$ when $\mathit{Req}$ is chosen by a party the principal does not trust.
\end{proposition}
\noindent The x402 facilitator compares the authorization's recipient and value with the merchant's \texttt{payTo} and amount (\texttt{facilitator.py}, lines 228--262). A merchant that sets \texttt{payTo} to a look-alike address, quotes a value with the wrong decimals or selects another network obtains a verified authorization. The mandate never reaches the verifier.

\begin{proposition}[Simulation]
Simulating the intended execution of $\sigma$ decides membership in $\sem{\M}$ of one sequence $s\in\Auth_W(\sigma)$. If $s\in\sem{\M}$ and $\Auth_W(\sigma)$ also contains a sequence outside $\sem{\M}$ that differs from $s$ in executor, recipient, time or repetition, simulation accepts an object that does not refine $\M$.
\end{proposition}
\noindent An allowance-creating object changes no balance when simulated, so simulation sees none of its authority.

\section{\sys{}: Computing Authority Footprints}
\label{sec:closure}
\sys{} maps a wallet object to a footprint $F=(c,a,R,A,[f,l],e)$: chain, asset, recipient set $R$, cumulative amount bound $A\le\infty$, window and executor $e$, the party that can submit or redeem the object. Each of $a$, $R$ and $e$ is either concrete or $\top$, meaning ``chosen by whoever executes''. A footprint with no asset ($a=\bot$) grants no token authority. The concretization $\gamma(F)$ is the set of transfer sequences on $(c,a)$ with recipients in $R$, times in $[f,l]$ and total at most $A$. \sys{} accepts iff $a=\bot$ or
\[
c=c_\M,\; a=a_\M,\; R\subseteq Y,\; A\le\kappa,\; [f,l]\subseteq[t_0,t_1].
\]
Each failed conjunct is reported as a violated dimension, with one convention: a footprint on another chain reports \emph{chain} and not also \emph{asset}, and the amount is compared only when the asset matches or is $\top$. Since $\gamma(F)\subseteq\sem{\M}$ whenever the conditions hold, \sys{} is sound if $\Auth_W(\sigma)\subseteq\gamma(F(\sigma))$.

\begin{table*}[t]
\centering
\caption{Footprints computed by \sys{}. $\top$: chosen by the executor; $\infty$: unbounded; $s$: signing time; $\bot$: no token authority. Recipients, amounts and windows refer to the signed fields named in each row. ``Registered'' means the address is in \sys{}'s registry of trusted contract code.}
\label{tab:footprint}
\small
\begin{tabularx}{\textwidth}{@{}>{\raggedright\arraybackslash}p{3.75cm}>{\raggedright\arraybackslash}p{1.65cm}>{\raggedright\arraybackslash}p{2.45cm}>{\raggedright\arraybackslash}p{3.7cm}>{\raggedright\arraybackslash}X@{}}
\toprule
Family (signed object) & Recipients $R$ & Amount $A$ & Window $[f,l]$ & Reason (consuming code)\\
\midrule
1 ERC-20 \texttt{transfer} (tx) & $\{\mathit{to}\}$ & value & $[s,s]$ & immediate execution in the model\\
2 ERC-20 \texttt{approve} (tx) & $\top$ & value ($\infty$ if max) & $[s,\infty)$ & spender calls \texttt{transferFrom} to any recipient until revoked\\
3 EIP-2612 permit (token domain) & $\top$ & value ($\infty$ if max) & $[s,\infty)$ & the deadline bounds submission of the permit, not spending\\
4 ERC-3009 transfer (token domain) & $\{\mathit{to}\}$ & value & $[\mathit{after}{+}1,\mathit{before}{-}1]$ & strict bounds; nonce single use; any submitter\\
5 ERC-3009 receive (token domain) & $\{\mathit{to}\}$ & value & as family 4 & the submitter must be \texttt{to}\\
6 Permit2 signature transfer & $\top$ & permitted & $[s,\mathit{deadline}]$ & the spender supplies the recipient; nonce single use\\
7 x402 exact (Permit2 witness) & $\{\mathit{witness.to}\}$ & permitted & $[\mathit{validAfter},\mathit{deadline}]$ & registered proxy pays the witnessed recipient\\
8 x402 upto (Permit2 witness) & $\{\mathit{witness.to}\}$ & permitted (max) & $[\mathit{validAfter},\mathit{deadline}]$ & only the named facilitator settles, any amount up to the maximum\\
9 Permit2 allowance & $\top$ & amount ($\infty$ if $2^{160}{-}1$) & $[s,\mathit{exp}]$; $[s,\mathit{sigDeadline}]$ if $\mathit{exp}{=}0$ & allowance spendable until expiration\\
10 ERC-7710 delegation & \multicolumn{3}{>{\raggedright\arraybackslash}p{7.9cm}}{composed from caveats (this section)} & caveats narrow unrestricted authority\\
11 signed \texttt{transfer} held by agent & $\{\mathit{to}\}$ & value & $[s,\infty)$ & valid until its nonce is used; anyone can broadcast it\\
12 EIP-7702 authorization & $\bot$ if registered, else $\top$ & $0$, else $\infty$ & $[s,\infty)$ if unregistered & installed code acts for the account\\
13 pair of family 4 authorizations & union & sum & hull & distinct nonces, each redeemable once\\
2, 3 with spender Permit2 & \multicolumn{3}{>{\raggedright\arraybackslash}p{7.9cm}}{$\bot$: no token authority without a further Permit2 signature} & Permit2 code\\
Witness to unregistered spender & $\top$ & permitted & $[s,\mathit{deadline}]$ & fail closed\\
\bottomrule
\end{tabularx}
\end{table*}

\paragraph{Spender-dependent semantics.} For Permit2 witness transfers the signed spender is the contract allowed to consume the signature, and the witness means only what that contract does with it. Permit2 checks that the signature covers the witness hash, but the transfer's recipient is an argument chosen by the spender. The x402 Exact proxy passes \texttt{witness.to} as that argument and transfers the full permitted amount; the Upto proxy additionally requires \texttt{msg.sender} to equal \texttt{witness.facilitator} and lets it choose any amount up to the permitted one. A contract that accepts the same witness type and ignores \texttt{witness.to} obtains a valid transfer to any recipient. \sys{} therefore holds a registry of settlement contracts whose code it trusts; a witness signed to an unregistered spender receives $R=\top$, as does any unknown verifying contract or calldata selector.

\paragraph{Delegations.} An ERC-7710 delegation authorizes the delegate to make the principal's account execute calls. \sys{} starts from unrestricted authority (any target and method, $R=\top$, $A=\infty$, window $[s,\infty)$) and lets each caveat whose enforcer it knows narrow it. A transfer-amount caveat fixes the target token, restricts the method to \texttt{transfer} and bounds $A$. A periodic-transfer caveat does the same per period (\texttt{ERC20PeriodTransferEnforcer.sol}, lines 187--188); over a window it yields $A=p\cdot n$ for period amount $p$ and the $n$ periods that intersect the window, and $A=\infty$ for an unbounded window. Target and method caveats intersect the allowed sets. An allowed-calldata caveat that pins the first argument binds $R$ only if the method set is $\{\texttt{transfer}\}$: if \texttt{approve} remains callable, the pinned address receives an allowance that it can spend anywhere. A timestamp caveat bounds the window with strict inequalities. If more than one target, or no known target, remains, the asset is $\top$. Unknown caveats are ignored, which is sound because a caveat can only restrict what the delegate may execute.

\paragraph{Objects that only enable registered code.} An EIP-7702 authorization that installs a registered delegator, whose code executes calls only for the delegation manager, grants no authority by itself: tokens move only under a separately signed delegation, which \sys{} checks on its own. An approval or EIP-2612 permit whose spender is Permit2 is analogous, because Permit2 moves tokens only under its own signatures. \sys{} gives both the footprint $\bot$. An authorization that installs any other code receives unrestricted authority over every asset.

\paragraph{Transaction timing.} The family 1 footprint in Table~\ref{tab:footprint} assumes that if the transfer executes, it does so at signing time~$s$. In Ethereum, a broadcast transaction can remain pending before a block includes it~\cite{spec:ethereum-transactions}. An ordinary ERC-20 transfer has no expiry, so without the immediate-execution assumption its conservative window is $[s,\infty)$, as in family 11.

\paragraph{Soundness.}
\begin{theorem}[Soundness]
\label{thm:sound}
Assume the registered contracts implement the semantics cited in Table~\ref{tab:footprint} and in this section, that the principal's standing state $W$ contains no authorizations other than approvals of Permit2 and registered delegator code, and that every family 1 transfer that executes does so at signing time~$s$. For family 13, let $\sigma$ consist of two family 4 authorizations from the same principal on the same chain and token, with distinct nonces. Then for every $\sigma$ in Table~\ref{tab:footprint}, $\Auth_W(\sigma)\subseteq\gamma(F(\sigma))$. Hence \sys{} accepting $\sigma$ implies $\sigma\sqsubseteq\M$.
\end{theorem}
\noindent\emph{Proof sketch.} We consider every entry point through which $\sigma$ can be consumed. Families 1 and 11 are consumed by including the signed transaction, at most once because of its nonce. For family 1, the timing assumption gives the window $[s,s]$. For family 11, the nonce bounds the number of executions and leaves the latest possible execution time unbounded. Families 4 and 5 are consumed by \texttt{transferWithAuthorization} or \texttt{receiveWithAuthorization}, at most once because of the authorization nonce, within the strict timestamp bounds. Family 7 is consumed by the Exact proxy's \texttt{settle} or \texttt{settleWithPermit}; both pass \texttt{witness.to} and the permitted amount to Permit2, whose nonce bitmap makes the signature single-use across the single and batch entry points. Family 8 adds the facilitator check and a caller-chosen amount at most the permitted one. For family 6 the signature is single-use, so the total is the permitted amount, but the recipient is a caller argument. For families 2, 3 and 9 the allowance is spent through \texttt{transferFrom} (or Permit2's single and batch \texttt{transferFrom}) with a caller-chosen recipient; each spend decrements the allowance, except at the maximal values, which both contracts treat as unlimited. ERC-20 allowances never expire; Permit2 allowances expire at \textit{exp}, or in the block of submission if $\mathit{exp}=0$. For family 10, the delegation manager executes a call only if every caveat enforcer's hooks succeed, so every executed call lies in the intersection of the call sets the enforcers admit. Each rule above over-approximates its enforcer's set; the amount and periodic enforcers keep state across redemptions, so the bound holds for the whole sequence of redemptions. When the method set is $\{\texttt{transfer}\}$ and the target is the token, each executed call causes exactly one transfer to the pinned recipient. For family 12 a registered delegator executes only calls the delegation manager forwards, which are covered by family 10. Objects with spender Permit2 enable only Permit2 code paths, each of which requires a further principal signature or a Permit2 allowance. For family 13, label each debit by the authorization nonce that the token consumes. Each successful redemption transfers its signed value $v_i$ to its signed recipient within its own window and marks that nonce used. Neither authorization creates an allowance or changes the fields of the other. In any order of submissions, the total is therefore $\sum_{i\in I}v_i\le v_1+v_2$, where $I\subseteq\{1,2\}$ indexes the redeemed nonces. Every recipient belongs to the union and every execution time lies in the hull of the two windows. \qed

All 90 family 13 entries in the corpus satisfy these pair conditions. A group in which one signature enables another requires a separate composition argument. The theorem relies on the registry and the cited semantics. Section~\ref{sec:results} tests both against the contracts and tests the checker's rules by mutation.

\paragraph{What can refine.} The definition implies that some families can never refine a payee-bound mandate, whatever their fields say.

\begin{theorem}[Name, cap, expire]
\label{thm:char}
Under the assumptions of Theorem~\ref{thm:sound}, let $\M$ have a finite payee set $Y$, a finite cap and a finite expiry $t_1$, and assume the principal's balance exceeds $\kappa$. Under the strict definition, an object $\sigma$ of families 1--13 with a footprint other than $\bot$ refines $\M$ only if (i) its recipients are fixed by signed fields or registered code, (ii) its cumulative amount is bounded by signed fields, and (iii) its feasible execution times lie within the mandate window. This bound comes from the signed fields and consuming code, or from the immediate-execution assumption for family 1. The full acceptance condition of \sys{}, including chain and asset, is sufficient by Theorem~\ref{thm:sound}.
\end{theorem}
\noindent\emph{Proof sketch.} Necessity: if (i) fails, the executor chooses a recipient outside $Y$; if (ii) fails, it repeats or enlarges transfers past $\kappa$; if (iii) fails, a transfer executes before $t_0$ or after $t_1$. Each yields a sequence in $\Auth_W(\sigma)\setminus\sem{\M}$. Sufficiency is Theorem~\ref{thm:sound} with the acceptance condition of \sys{}. \qed

\begin{corollary}
Under the strict definition, ERC-20 approvals, EIP-2612 permits, Permit2 allowances and Permit2 signature transfers without a registered witness never refine a payee-bound mandate, and a signed transaction held by another party never refines a mandate with an expiry. Under the relaxed definition, Permit2 signature transfers and Permit2 allowances whose spender is a payee can refine; ERC-20 allowances still cannot, because they do not expire.
\end{corollary}

\paragraph{Implementation.} \sys{} is about 230 lines of dependency-free Python. It consumes exactly what a wallet receives: the typed data of \texttt{eth\_signTypedData\_v4}, the chain, target and calldata of a transaction, or an EIP-7702 authorization. It never sees the oracle's role labels.

\section{Deployment}
\label{sec:deploy}
\paragraph{Where the mandate comes from.} \sys{} needs the translated mandate $(c,a,Y,\kappa,[t_0,t_1])$. The principal's trusted surface already shows the merchant, the amount and the expiry when the open mandate is approved. The wallet that renders this surface resolves the merchant to payout addresses, the currency to a token contract and chain, and the amount to base units, using registries it trusts, and stores the result with the session that the agent uses. In an ERC-7715 flow the requested permission itself carries these fields and is approved on the wallet's own surface.

\paragraph{Where \sys{} runs.} The natural place is the wallet's signing path: \texttt{eth\_signTypedData\_v4}, \texttt{eth\_sendTransaction} and EIP-7702 authorization requests that arrive during an agent session can be checked against the active mandate before the key is used. For a mandate with a finite deadline, a deployed wallet should refuse ordinary transfers with no execution-time expiry and use a payment path whose consuming contract enforces the expiry. A wallet that refuses an object names the violated dimensions, which tells the agent what to fix. A facilitator can run the same check if the client sends a commitment to the mandate together with the payment, which would turn x402 verification from requirement-relative into mandate-relative.

\paragraph{Registry maintenance.} The registry lists contracts whose consuming code \sys{} trusts: Permit2, settlement proxies, delegation managers, caveat enforcers and EIP-7702 delegators, with one footprint rule each. These are a few audited contracts per chain that change rarely, and they are the same contracts a clear-signing format such as ERC-7730~\cite{spec:erc-7730} already describes per contract. An object that touches anything outside the registry is rejected, so a stale registry costs usability and never safety.

\section{Methodology}
\label{sec:method}
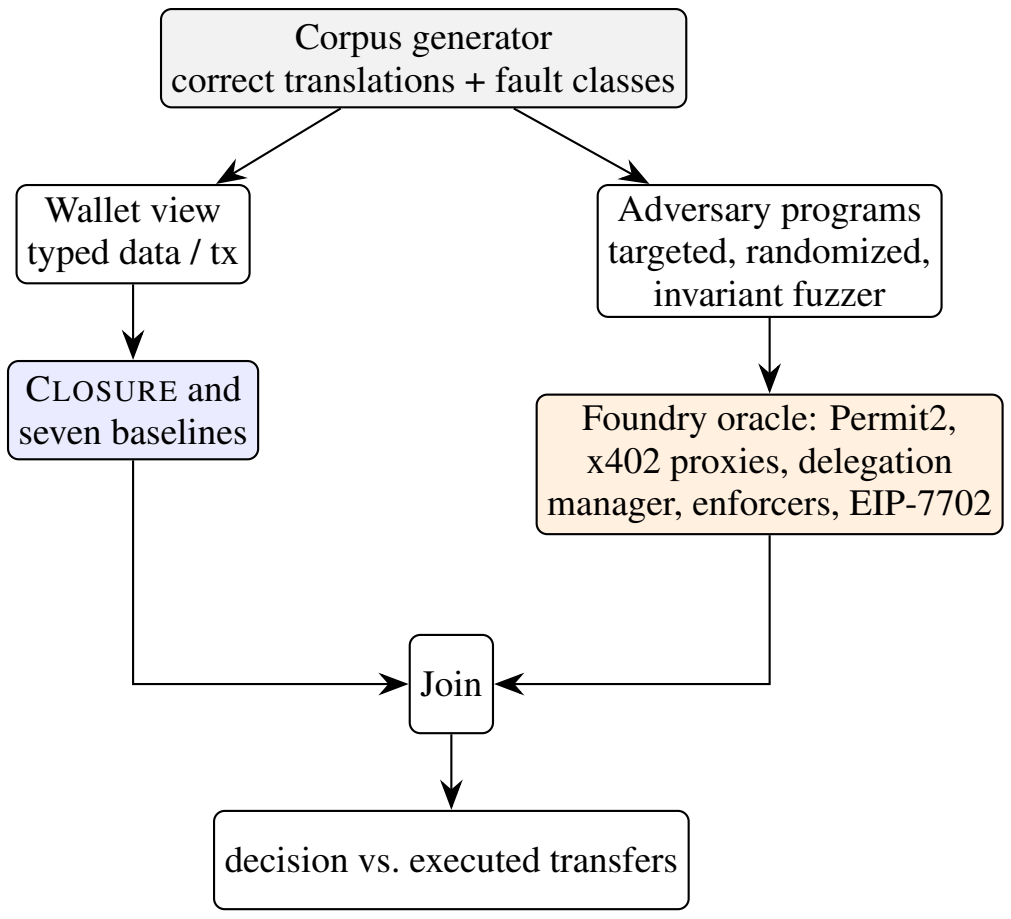
\begin{figure}[t]
\centering
\begin{tikzpicture}[font=\scriptsize, node distance=3mm and 4mm,
  box/.style={draw, rounded corners=2pt, align=center, minimum height=6.5mm, inner sep=2pt},
  arr/.style={-{Stealth[length=2mm]}}]
\node[box, fill=gray!10] (gen) {Corpus generator\\correct translations + fault classes};
\node[box, below left=5mm and -6mm of gen] (view) {Wallet view\\typed data / tx};
\node[box, below right=5mm and -6mm of gen] (prog) {Adversary programs\\targeted, randomized,\\invariant fuzzer};
\node[box, below=5mm of view, fill=blue!8] (clo) {\sys{} and\\seven baselines};
\node[box, below=5mm of prog, fill=orange!12] (orc) {Foundry oracle: Permit2,\\x402 proxies, delegation\\manager, enforcers, EIP-7702};
\node[box, below=9mm of $(clo.south)!0.5!(orc.south)$] (j) {Join};
\node[box, below=5mm of j] (out) {decision vs.\ executed transfers};
\draw[arr] (gen) -- (view); \draw[arr] (gen) -- (prog);
\draw[arr] (view) -- (clo); \draw[arr] (prog) -- (orc);
\draw[arr] (clo.south) |- (j.west); \draw[arr] (orc.south) |- (j.east); \draw[arr] (j) -- (out);
\end{tikzpicture}
\caption{Evaluation pipeline. Checkers see only the wallet view; ground truth comes only from executed transfers.}
\label{fig:pipeline}
\end{figure}

\paragraph{Executable oracle.} The oracle is a Foundry test harness (Figure~\ref{fig:pipeline}). It deploys Permit2 from Uniswap's source (commit \texttt{cc56ad0}), the x402 Exact and Upto Permit2 proxies from the x402 repository (commit \texttt{0e6e8c5}), the MetaMask \texttt{DelegationManager} and its transfer-amount, periodic-transfer, allowed-calldata, allowed-targets, allowed-methods, timestamp and limited-calls enforcers (commit \texttt{b4b8c41}), and four ERC-20 tokens with EIP-2612 and ERC-3009 (USDC and EURC with six decimals, a WETH-like token with eighteen, and USDC under a second chain identifier). Section~\ref{sec:limitations} lists the substitutions. For delegations the principal's address is upgraded with an EIP-7702 designator to a stateless delegator that executes single calls for the delegation manager and validates ERC-1271 signatures by the address's own key. Two further contracts model adversarial code: a sweeper that an unregistered EIP-7702 authorization installs, and a look-alike settlement contract that accepts the x402 exact witness type and ignores \texttt{witness.to}. Each typed-data artifact is signed with the principal's key exactly as the wallet view states; transactions and EIP-7702 designations are executed with Foundry cheatcodes (\texttt{prank}, \texttt{etch}) that have the same on-chain effect without a signature. Every program starts from the same snapshot, in which the principal has approved Permit2, as Permit2 users do. The oracle records every nonzero \texttt{Transfer} event out of the principal with its token, recipient, value, block time and the operation that caused it. Contract spenders act only through their public functions.

\paragraph{Adversary programs.} A program is a sequence of up to eight operations: advance time, submit the artifact as a chosen caller with a chosen recipient and amount, spend a resulting allowance, or probe residual allowances. Targeted programs submit at the earliest valid time, one second after the mandate expires, thirty days later and one second before the signed deadline; submit early and spend late or exactly at an inclusive expiration; replay; route a delegation through \texttt{approve}; redeem a periodic delegation once in every period the window touches; send to the recipient pinned by a caveat; and drain the realistic wallet balance. Each artifact also receives 20 randomized programs over callers, recipients, amounts and times. The harness executes wallet-broadcast transactions (families 1 and 2) once at signing time. The adversary controls subsequent actions. An artifact is \emph{unauthorized} iff some program realizes a transfer sequence outside $\sem{\M}$, and the violated dimensions are recorded.

\paragraph{Independent adversaries.} We wrote the program library knowing the footprint rules, so a shared blind spot could make the checker and the oracle agree for the wrong reason. Two checks do not depend on that library, although the fuzzer reuses the harness's submit and spend operations. Forge's invariant fuzzer drives one artifact per pair of class and family with its own sequences of submissions, spends and time jumps (\InvSample{} artifacts, 64 runs of depth 200 each) and asserts after every call that no transfer has left the mandate. Mutation analysis weakens one rule of \sys{} at a time (\MutTotal{} mutants, such as an inclusive \texttt{validBefore}, a permit deadline that ends spending, a periodic amount read as a total cap, or trusting any EIP-7702 code) and asks whether the corpus detects the change.

\paragraph{Balances.} We run every program twice. In the authority run the principal holds $2^{200}$ base units of each token, so that execution is limited by authority and never by funds. In the exposure run the principal holds 2,000 USDC, 2,000 USDC on the second chain, 2,000 EURC and one WETH, and we value out-of-mandate transfers at fixed illustrative prices (EURC 1.08 dollars, WETH 3,000 dollars).

\paragraph{Corpus.} Each artifact has a mandate with a cap between 1 and 500 dollars and an expiry between one hour and seven days. Correct translations use families 1, 4, 5, 7, 8 and 10. Two benign classes test exact reasoning: \emph{benign boundary} artifacts sit exactly on the mandate's edges (first valid second equal to $t_0$, last equal to $t_1$, a periodic allowance whose total over the window is at most the cap, an x402 upto payment to a second payee), and \emph{Permit2 onboarding} artifacts are the gas-sponsoring approvals and permits of Table~\ref{tab:paths}. The remaining classes model mistaken or manipulated translations: a look-alike payee that shares the first and last four hexadecimal digits; amounts computed with eighteen decimals or with cents read as dollars; another asset or chain; a deadline beyond the mandate; \emph{authority escalation}, an allowance or signature transfer whose displayed payee, amount, asset, chain and expiry all match the mandate; unlimited allowances; a Permit2 signature transfer without witness; an upto maximum above the cart; swapped recipient and amount arguments; a substituted spender or facilitator; four delegation classes (recipient not pinned; unscoped, with only an amount caveat or with a recipient pin that leaves the method open; periodic allowance; no amount caveat); \emph{near misses} one base unit or one second outside the mandate, including a payee-spender Permit2 allowance usable one second after $t_1$; authorizations valid before $t_0$; a signed transaction held by the agent; an EIP-7702 authorization of unregistered code; a pair of ERC-3009 authorizations for one payment, as a retry or split produces; and a witness signed to a look-alike settlement contract. Thirty artifacts are generated per class and applicable family, from a fixed seed.

\paragraph{Baselines.} \emph{Simulation} accepts if the intended execution, by the expected executor at the earliest valid time for the cart amount, stays within $\sem{\M}$ and, for objects that should pay, moves value. \emph{Simulation + state} additionally rejects any residual allowance after the intended execution, any allowance-creating object and any EIP-7702 authorization. \emph{Fact binding} accepts if the displayed payee (the recipient, else the spender or delegate), amount, asset and chain match the mandate. \emph{Fact binding + expiry} additionally requires a displayed deadline, expiration or timestamp caveat at most $t_1$, and rejects objects that display none. \emph{Policy guard} models a transaction guard such as Safe's~\cite{safeguards}: every displayed address is a payee or registered infrastructure (Permit2, settlement proxies, the facilitator, the principal's agent), no allowance is unlimited, and the amount, asset and chain match. \emph{Type allow-list} is fact binding with expiry restricted to the families that can refine under Theorem~\ref{thm:char}; it admits registered EIP-7702 authorizations without a fact check. \emph{x402 verify} transcribes the exact-scheme checks of the x402 Python facilitator (\texttt{facilitator.py}, lines 188--262, and the canonical-spender check of \texttt{permit2\_utils.py}, lines 205--212) and is evaluated only on the artifacts to which it applies (single ERC-3009 \texttt{transferWithAuthorization} objects and single Permit2 witnesses, families 4, 7 and 8), against honest requirements derived from the mandate and against merchant requirements that copy \texttt{payTo}, amount, asset and network from the artifact itself, as a merchant who controls the requirements would.

\section{Results}
\label{sec:results}
The final run executed \NPrograms{} programs (\NFuzz{} randomized) over \NCases{} artifacts. Execution showed \NViol{} artifacts to be unauthorized and \NBenign{} to be authorized.

\begin{table*}[t]
\centering
\caption{Unauthorized artifacts accepted, per class (benign classes show ``--''). Families as in Table~\ref{tab:footprint}. ``Unauth.'' counts artifacts for which some adversary program realized a transfer outside the mandate. x402 uses honest requirements and counts only artifacts of families 4, 7 and 8 (n/a for classes with none). Loss: out-of-mandate value realized in the exposure run (Section~\ref{sec:method}), summed over the class, in dollars.}
\label{tab:classes}
\footnotesize
\setlength{\tabcolsep}{4pt}
\begin{tabular}{@{}llrrrrrrrrrr@{}}
\toprule
Class & Families & $n$ & Unauth. & Sim. & Sim.+st. & x402 & Fact+exp. & Policy & Type & \sys{} & Loss \\
\midrule
benign & 1, 4, 5, 7, 8, 10 & 180 & 0 & -- & -- & -- & -- & -- & -- & -- & 0 \\
payee lookalike & 1, 4, 5, 7, 8, 10 & 180 & 180 & 0 & 0 & 0 & 0 & 0 & 0 & 0 & 46,751 \\
decimals 18 & 1, 4, 7, 8 & 120 & 120 & 30 & 30 & 0 & 0 & 0 & 0 & 0 & 52,672 \\
minor units & 1, 4, 7, 8 & 120 & 120 & 30 & 30 & 0 & 0 & 0 & 0 & 0 & 51,345 \\
asset substitution & 1, 4, 7, 8 & 120 & 120 & 0 & 0 & 0 & 0 & 0 & 0 & 0 & 44,562 \\
chain substitution & 1, 4, 7, 8 & 120 & 120 & 0 & 0 & 0 & 0 & 0 & 0 & 0 & 29,861 \\
window overrun & 4, 5, 7, 8, 10 & 150 & 150 & 150 & 150 & 90 & 0 & 150 & 0 & 0 & 39,134 \\
authority escalation & 2, 3, 6, 9 & 120 & 120 & 120 & 30 & n/a & 90 & 120 & 0 & 0 & 31,426 \\
unlimited allowance & 2, 3, 9 & 90 & 90 & 90 & 0 & n/a & 0 & 0 & 0 & 0 & 180,000 \\
recipient unbound & 6 & 30 & 30 & 30 & 30 & n/a & 0 & 30 & 0 & 0 & 8,204 \\
upto overcap & 8 & 30 & 30 & 30 & 30 & 0 & 0 & 0 & 0 & 0 & 34,419 \\
field swap & 1, 4 & 60 & 60 & 0 & 0 & 0 & 0 & 0 & 0 & 0 & 0 \\
executor substitution & 6, 8 & 60 & 30 & 30 & 30 & n/a & 0 & 0 & 0 & 0 & 5,233 \\
delegation recipient unbound & 10 & 30 & 30 & 30 & 30 & n/a & 0 & 30 & 0 & 0 & 7,134 \\
delegation unscoped & 10 & 30 & 30 & 30 & 30 & n/a & 0 & 16 & 0 & 0 & 10,032 \\
delegation periodic & 10 & 30 & 30 & 30 & 30 & n/a & 0 & 30 & 0 & 0 & 25,501 \\
delegation amount unbounded & 10 & 30 & 30 & 30 & 30 & n/a & 0 & 0 & 0 & 0 & 16,576 \\
benign boundary & 4, 7, 8, 10, 12, 13 & 180 & 0 & -- & -- & -- & -- & -- & -- & -- & 0 \\
near miss & 4, 7, 8, 9, 10, 13 & 180 & 180 & 120 & 90 & 30 & 55 & 120 & 55 & 0 & 14,121 \\
early start & 4, 7, 10 & 90 & 90 & 0 & 0 & 0 & 90 & 90 & 90 & 0 & 23,376 \\
held transaction & 11 & 30 & 30 & 30 & 30 & n/a & 0 & 30 & 0 & 0 & 7,896 \\
code authorization & 12 & 30 & 30 & 0 & 0 & n/a & 0 & 0 & 0 & 0 & 48,356 \\
retry split & 13 & 30 & 30 & 30 & 30 & n/a & 30 & 30 & 30 & 0 & 8,106 \\
settlement substitution & 7 & 30 & 30 & 30 & 30 & 0 & 30 & 0 & 0 & 0 & 7,649 \\
permit2 onboarding & 2, 3 & 60 & 0 & -- & -- & -- & -- & -- & -- & -- & 0 \\
\bottomrule
\end{tabular}

\end{table*}

\begin{figure*}[t]
\centering
\includegraphics[width=0.86\textwidth]{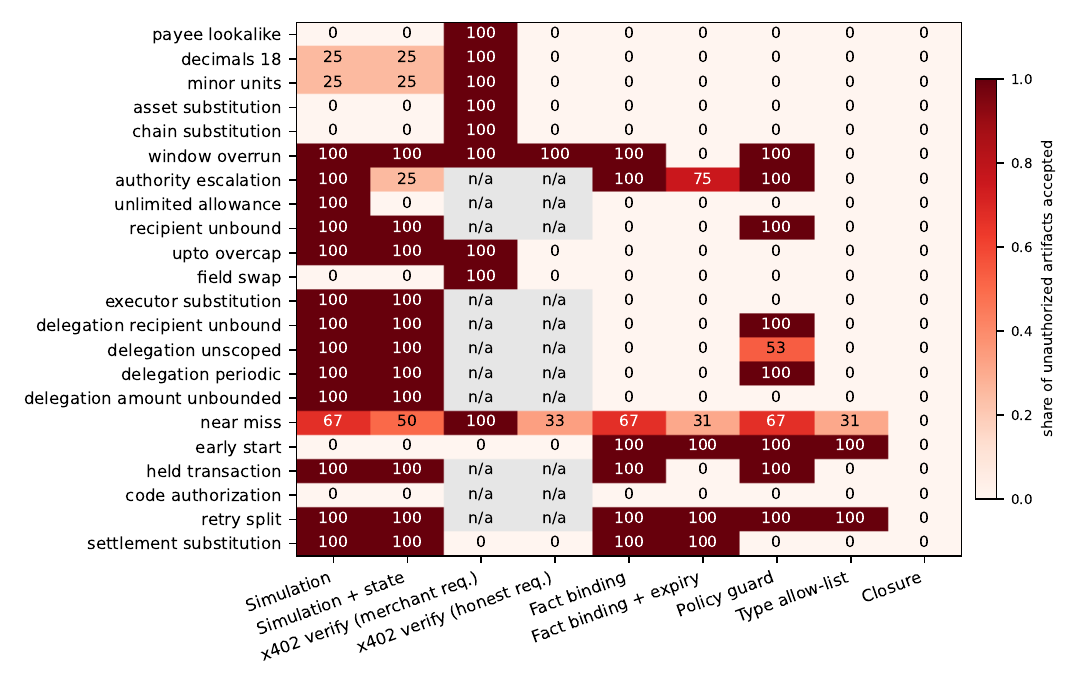}
\caption{Share of unauthorized artifacts accepted by each check, per class (percent; darker is worse; n/a where x402 verification does not apply).}
\label{fig:acceptance}
\end{figure*}

\paragraph{RQ1: deployed checks admit unauthorized artifacts.} Table~\ref{tab:classes} and Figure~\ref{fig:acceptance} report acceptance per class. Every baseline admits at least one class completely. Simulation accepts \SimFA{} unauthorized artifacts: window overruns, allowance-creating objects, unwitnessed Permit2 transfers, unsafe delegations, held transactions, duplicated authorizations and inflated upto maxima, because in each case the intended execution stays within the mandate or moves no balance. Checking for residual allowances lowers this to \SimStateFA{}, but it cannot see a deadline beyond the mandate, a delegation that stays valid after its intended use, or a held transaction that is still broadcastable. Where it applies, x402 verification with honest requirements catches look-alike payees, decimal errors and asset or chain substitutions, and its canonical-spender check catches look-alike settlement contracts; it accepts \XHonFAApp{} of \XApp{} unauthorized artifacts, because it compares the deadline with the current time and never with a mandate. With requirements written by the merchant it accepts \XMerFAApp{}. Fact binding accepts \FactFA{} unauthorized artifacts; adding the expiry check leaves \FactTFA{}. The policy guard accepts \PolicyFA{} and wrongly rejects \PolicyFR{} authorized artifacts. The type allow-list accepts \TypeFA{}: it admits the families that \emph{can} refine (minus ERC-20 transfers, which display no expiry), but whether a given object does depends on details such as an exclusive \texttt{validBefore}, a start before $t_0$, a periodic allowance that accumulates over the window, or a retry that duplicates an authorization.

The \FBUnauth{} artifacts accepted by both fact binding with expiry and the policy guard are the core finding. Ninety are EIP-2612 permits, Permit2 signature transfers and Permit2 allowances naming the payee as spender. Each displays the right payee, amount, token, chain and an in-mandate expiry, and in execution each lets an adversary move the mandate amount to its own address, after the mandate has expired, or both. By Theorem~\ref{thm:char} no field-level check can repair these families; only refusing them can. (An \texttt{approve} transaction displays no expiry and fails the expiry check.) The other 175 (90 early starts, 55 near misses and 30 retry pairs) pass field checks because each field is individually within the mandate: ERC-3009, x402 and delegation authorizations that become valid before $t_0$, periodic delegations and ERC-3009 pairs whose total exceeds the cap by a few base units, and retry pairs that authorize the same payment twice.

\paragraph{A worked trace.} Artifact \TraceCase{} is an EIP-2612 permit naming the payee as spender for exactly the mandate cap of \TraceValue{} USDC, with a deadline \TraceDeadlineH{} hours after signing; the mandate expires after \TraceExpiryH{} hours. Fact binding with expiry, the policy guard and simulation all accept it. The program that submits the permit at once and spends the allowance later moved \TraceValue{} USDC to the adversary's address \TraceDays{} days after signing, 30 days after the mandate expired. \sys{} reports \emph{recipient} and \emph{time}, the dimensions that execution violated.

\paragraph{RQ2: \sys{} agrees with execution.} Under the timing model of Section~\ref{sec:method}, \sys{} accepted \ClosureFA{} of \NViol{} unauthorized artifacts and rejected \ClosureFR{} of \NBenign{} authorized ones. For all \ReasonExact{} unauthorized artifacts its reported dimensions equal the dimensions execution violated. Under the relaxed definition \RelViol{} artifacts are unauthorized (\StrictOnly{} artifacts are unauthorized only under the strict one); relaxed \sys{} accepts \RelFA{} of them and rejects \RelFR{} relaxed-authorized artifacts. \NProgAccepted{} programs ran against the artifacts \sys{} accepted and none realized a transfer outside the mandate, the empirical counterpart of Theorem~\ref{thm:sound} on the contracts. Every rejection is witnessed by a concrete violating execution: targeted programs found a violation for \TargetedFound{} unauthorized artifacts, and the randomized programs found none that the targeted ones missed.

\paragraph{Independent adversaries.} The invariant fuzzer, which composes its own operation sequences, found a transfer outside the mandate for \InvFound{} of the \InvUnauth{} unauthorized artifacts in its sample of \InvSample{}, and none on any authorized or \sys{}-accepted artifact. It missed \InvMissed{} unauthorized artifacts. Five need a precise time or amount: acting in a one-second window at the mandate's edge or before $t_0$, or settling just above the cap. The other two need a specific sequence: settling an upto authorization after the mandate expires through the named facilitator, and spending a delegation that lacks an amount caveat beyond the cap. Of the \MutTotal{} mutants of \sys{}, \MutDecision{} change at least one accept or reject decision against execution and \MutAny{} are detected when reported dimensions and the relaxed variant are also compared (Appendix~\ref{app:mutants}). Several rules change only reported dimensions or relaxed decisions, because another dimension already rejects the object under the strict definition; for example, every amount-bounding enforcer already restricts the method to \texttt{transfer}. The undetected mutant reads an approval of $2^{256}{-}1$ as finite, which is equivalent: that value exceeds every cap either way.

\paragraph{Reference-client checkouts.} We placed \sys{} in the signing callback of the x402 TypeScript client (v2.31.0) and used the reference facilitator to settle purchases on Anvil. Each route used a fresh local chain and Circle's FiatTokenV2\_2 implementation compiled with solc 0.6.12. The buyer authorized up to 25 USDC to one shop per checkout. On the direct ERC-3009 route, the 25 USDC purchase settled and a fresh request for the same price to another payee reached \sys{}, which refused it before signing.

On the Permit2 route, the client's signature names a settlement proxy as spender and places the shop's address in the signed witness. We installed the pinned Permit2 and x402 Exact proxy runtime code at the addresses expected by the client, and gave Permit2 a 100 USDC allowance. The 25 USDC checkout returned HTTP 200. Its receipt targeted the proxy and contained one token transfer from the buyer to the approved shop for exactly 25 USDC. We then issued two fresh quotes: 25 USDC to another payee and 26 USDC to the approved shop. The client limit was 40 USDC and the buyer still held 75 USDC. Both quotes reached the signing callback. \sys{} refused the first for its recipient and the second for its amount. Each refusal produced zero payment signatures and zero settlements, with no change to the three token balances or the remaining allowance. These observations cover the two tested routes and are reported separately from the generated corpus.

\paragraph{RQ3: exposure.} In the exposure run, \RealLossCases{} unauthorized artifacts let the programs move value outside the mandate. Split by artifact, \RealWithin{} dollars came from artifacts whose violations were only recipient or time, and \RealBeyond{} dollars from artifacts that also exceeded the cap or moved the wrong asset or chain. Inflated amounts behave differently across families. An exact-amount artifact whose value is $10^{12}$ times too large reverts for lack of funds and realizes nothing; the same error in an upto maximum, an allowance or a delegation lets the adversary take the whole balance. The signed amount does not determine the exposure; the authority does.

\paragraph{RQ4: cost and compatibility.} \sys{}'s median decision time per artifact family ranges from \LatMin{} to \LatMax{}\,$\mu$s, with a 99th percentile of at most \LatPMax{}\,$\mu$s, on one laptop core. In the test harness, \sys{} rejected none of the \NBenign{} generated authorized artifacts, which cover every path of Table~\ref{tab:paths}: ERC-3009 authorizations, x402 exact and upto witnesses, ERC-7710 delegations that combine an amount or periodic caveat with a recipient pin and a timestamp caveat, registered EIP-7702 upgrades and the x402 Permit2-onboarding objects are all accepted when their fields are right. Fact binding, simulation with state and the type allow-list reject every Permit2-onboarding object, and the policy guard rejects the unlimited ones, so each would break x402's gas-sponsoring flows.

\section{Implications}
\label{sec:implications}
\paragraph{For agent commerce protocols.} A mandate that names a merchant identifier and a fiat amount cannot be checked against a wallet object. Protocols should carry, or let the principal's trusted surface resolve, the payee address set, the asset contract and chain, the cap in base units and an absolute expiry, and should bind the closed payment mandate to the hash of the exact wallet object. x402 verification should accept a mandate commitment from the client and check the authorization against it; requirements supplied by the resource server cannot serve that role. Payment paths should be restricted to families that name, cap and expire; x402's ERC-7710 method should verify the caveats of a delegation instead of simulating one redemption.

\paragraph{For wallets.} Wallets should compute and display authority instead of fields: ``anyone holding this can send up to 25 USDC to any address, forever'' is the truthful rendering of the permit in Section~\ref{sec:intro}. Clear-signing formats such as ERC-7730~\cite{spec:erc-7730} could carry the footprint rule of each registered contract. A wallet that serves agents should refuse objects whose footprint exceeds the active mandate and name the violated dimension, as \sys{} does.

\paragraph{For permission standards.} An ERC-7715 permission for an agent payment needs three caveats: an amount bound, a recipient pin with the method restricted to \texttt{transfer}, and a time bound. A recipient pin alone is not enough while \texttt{approve} remains callable. Periodic permissions accumulate: a daily allowance of $p$ over a seven-day mandate has authority $7p$, which a display of ``$p$ per day'' understates when the person approved a one-off cap. Deadline fields should be named for what they bound; the EIP-2612 deadline bounds submission and leaves the allowance unbounded.

\section{Related Work}
\label{sec:related}
\paragraph{Agent payment protocols.} Jiang et al.\ model x402, ACP, AP2 and MPP in Tamarin~\cite{jiang2026formal}. Aviv et al.\ analyze AP2~\cite{aviv2026mandate}, and Louck et al.\ show that an agent can manipulate a cart within the limits a user approved and propose binding the decision to the transaction~\cite{louck2026whisper}. Li et al.\ and Wang et al.\ analyze x402 and its deployments~\cite{li2026fiveattacks,wang2026x402}. Systematizations cover agentic commerce, agent-to-agent payments and the trust relation between agents and blockchains~\cite{mao2026sok,zhang2026sok,xia2026web4}. These works treat the payee as a merchant identifier or take the recipient from the merchant's requirements. We move the comparison to the wallet object and replace fact equality with refinement of the object's authority, checked against executed contract code. In Lowe's terms~\cite{lowe1997hierarchy}, fact binding gives agreement on data, while refinement constrains every future use of the signed object.

\paragraph{Agent security.} Prompt injection lets untrusted content steer tool-using agents~\cite{greshake2023indirect,debenedetti2024agentdojo,zhan2024injecagent}; Web3 agents can be steered through manipulated memory~\cite{patlan2025fakememories}. Defenses separate control and data flow or constrain capabilities~\cite{debenedetti2025camel,beurerkellner2025patterns}. PACE records policy decisions for DeFi agents~\cite{karanjai2026pace}. \sys{} is complementary: it bounds what any agent output can authorize, however the agent was steered.

\paragraph{Wallet and transaction security.} Payload-based phishing on Ethereum lures users into signing approvals and permits that hand over standing authority~\cite{ndss2025phishing}; address poisoning exploits signers who compare only the first and last digits of an address~\cite{tsuchiya2025poisoning}; the 2025 Bybit theft showed a signing interface displaying one transaction while the signers signed another~\cite{bybit2025blocksec}. Our authority-escalation and look-alike classes are the agent-payment counterparts of these attacks, with the difference that the displayed facts are correct. ERC-7730 and ERC-7739 make typed data readable~\cite{spec:erc-7730,spec:erc-7739}; readability shows fields, and \sys{} computes what the fields authorize. Smart-account guards enforce transaction policies~\cite{safeguards}, which our policy baseline models. The confused deputy~\cite{hardy1988confused} is the classical name for authority exercised on behalf of the wrong principal; allowances make every spender a potential deputy.

\section{Limitations}
\label{sec:limitations}
\emph{Contract substitutions.} The generated corpus uses a substitute token because solc 0.6.12 was unavailable during its offline run. The substitute implements ERC-3009 and EIP-2612 with the reference semantics of the ERCs on OpenZeppelin 5.4.0. Permit2's compiler pin (0.8.17) and the delegation framework's (0.8.23) were relaxed to compile with 0.8.28 under the upstream via-IR setting; Permit2's solmate transfer library was replaced by OpenZeppelin's SafeERC20; the delegation framework's ERC-7579 mode and execution libraries and its account-abstraction interfaces were replaced by shims that implement the reference single-call encoding; and the delegator is a minimal EIP-7702 stateless delegator, not MetaMask's. All other contract sources are upstream.
\emph{Corpus.} The artifacts are generated, not collected from deployments. The fault classes are chosen by us and do not estimate how often agents or merchants produce each one. Because the guarantee concerns the signed object, it does not depend on agent behavior, but prevalence remains unmeasured.
\emph{Ground truth.} The oracle's adversaries are a finite library, randomized programs and an invariant fuzzer, so the set of unauthorized artifacts is a lower bound. For accepted artifacts, absence of a violation is evidence and not proof; the proof is Theorem~\ref{thm:sound}, which depends on the registry and the family 1 timing assumption.
\emph{Model.} The harness applies transactions and EIP-7702 designations with Foundry cheatcodes. It models immediate execution for wallet-broadcast transactions and omits delays between broadcast and block inclusion. A second chain is modelled by a second token instance under a different chain identifier in one EVM. The model excludes the principal's other signatures, standing allowances other than to Permit2, and authority that does not move the principal's tokens.

\section{Conclusion}
\label{sec:conclusion}
At the boundary between agent commerce protocols and crypto wallets, the question that protects the person is what a signed object lets others do, and the facts it displays do not answer it. We defined that question as refinement of a mandate by the authority closure of a wallet object, showed that the checks in use today do not answer it, proved that compatible objects must constrain the recipient, total amount and execution time, and gave a checker that evaluates those bounds in microseconds under the stated model. Making the mandate a first-class input to wallets and facilitators would let the refusal happen where the signature is made.

{\sloppy\small
\bibliographystyle{plain}
\bibliography{references}}

\appendix
\section{Ethics Considerations}
\label{app:ethics}
All transfers executed on local EVMs with test tokens and ephemeral keys. The reference checkouts downloaded pinned dependencies and used local payment endpoints. No mainnet or testnet transaction was sent, no third-party system was probed, and no human subjects were involved. The weaknesses we describe are properties of standard interfaces and of how they are composed; the protocol behaviors we cite are documented in public specifications and source code. Before publication we will send the implications of Section~\ref{sec:implications} to the AP2, x402 and delegation-framework maintainers.

\section{Data and Code Availability}
\label{app:openscience}
The preprint source includes the manuscript, the reported results in its tables and figures, and the bibliography. The evaluation materials are maintained separately. They include \sys{}, the comparison checks, 12 offline tests, grouped counts by class and wallet-object family, and two reference-client checkouts with pinned dependencies and expected outcomes. The checkouts use disposable local accounts to verify signing decisions, settlement receipts and token balances. We restrict access to executable adversarial replay sequences and raw signing payloads because they can be adapted to reproduce unauthorized transfers.

\section{Validation History of the Harness}
\label{app:harness}
The two-sided design caught harness defects during development, each because \sys{} and the oracle disagreed. A replayed allowance permit was re-signed with the current nonce, so the oracle minted signatures the principal never gave. A delegation's caveat configuration was not encoded, so delegations ran without caveats. The adversary library never redeemed a delegation to the recipient pinned by its caveat, never spent a periodic allowance in every period, and let a held signed transaction execute twice. Zero-value \texttt{transferFrom} calls, which ERC-20 permits without any allowance, were counted as transfers. Every disagreement was resolved by inspecting execution traces against the contract source and fixing the harness; none was resolved by changing \sys{} to match the oracle.

\section{Ablation and Mutation Analysis}
\label{app:mutants}
\begin{table}[h]
\centering
\caption{Ablation. Each variant removes one component of \sys{}; errors are measured against execution on all \NCases{} artifacts.}
\label{tab:ablation}
\small
\begin{tabular}{@{}lrr@{}}
\toprule
Variant & False acc. & False rej. \\
\midrule
\sys{} & 0 & 0 \\
No settlement registry & 0 & 150 \\
No caveat semantics & 0 & 60 \\
No time dimension & 300 & 0 \\
No recipient dimension & 360 & 0 \\
No amount dimension & 434 & 0 \\
\bottomrule
\end{tabular}

\end{table}

Table~\ref{tab:ablation} removes one component at a time. Without the settlement-contract registry, \sys{} fails closed on every x402 Permit2 witness and wrongly rejects \AblNoRegFR{} authorized artifacts; without caveat semantics it rejects all \AblNoCavFR{} authorized delegations. Dropping the time, recipient or amount dimension admits \AblNoTimeFA{}, \AblNoRecFA{} and \AblNoAmtFA{} unauthorized artifacts.

\begin{table}[h]
\centering
\caption{Mutants of \sys{}. FA, FR: decisions that flip against execution; Reason: detected by decisions or reported dimensions; Relaxed: detected under the relaxed definition.}
\label{tab:mutants}
\scriptsize
\begin{tabular}{@{}lrrcc@{}}
\toprule
Mutant (rule weakened) & FA & FR & Reason & Relaxed \\
\midrule
3009-validBefore-inclusive & 0 & 30 & \checkmark & \checkmark \\
3009-validAfter-inclusive & 0 & 30 & \checkmark & \checkmark \\
permit-deadline-ends-spending & 0 & 0 & \checkmark & \checkmark \\
approve-binds-spender-as-recipient & 0 & 0 & \checkmark & -- \\
any-approval-is-standing & 60 & 0 & \checkmark & \checkmark \\
approve-max-is-finite & 0 & 0 & -- & -- \\
permit2-exp0-literal & 0 & 0 & \checkmark & \checkmark \\
permit2-allowance-to-deadline & 0 & 0 & \checkmark & \checkmark \\
ptf-spender-binds-recipient & 30 & 0 & \checkmark & -- \\
witness-trusts-any-spender & 30 & 0 & \checkmark & \checkmark \\
witness-ignores-validAfter & 0 & 30 & \checkmark & \checkmark \\
rawtx-expires-now & 30 & 0 & \checkmark & \checkmark \\
7702-any-code-trusted & 30 & 0 & \checkmark & \checkmark \\
period-is-total-cap & 25 & 0 & \checkmark & \checkmark \\
delegation-ignores-method-scope & 0 & 0 & \checkmark & -- \\
delegation-before-inclusive & 0 & 30 & \checkmark & \checkmark \\
delegation-ignores-after & 0 & 30 & \checkmark & \checkmark \\
batch-max-not-sum & 60 & 0 & \checkmark & \checkmark \\
cap-strict & 0 & 323 & \checkmark & \checkmark \\
ignore-notBefore & 90 & 0 & \checkmark & \checkmark \\
ignore-chain & 0 & 0 & \checkmark & -- \\
ignore-asset & 60 & 0 & \checkmark & \checkmark \\
\bottomrule
\end{tabular}

\end{table}

\end{document}